\documentclass[11pt,a4paper]{article}

\usepackage[
pdftitle={Viscoelastic-electromagnetism and Hall viscosity},pdfauthor={Yoshimasa Hidaka, Yuji Hirono, Taro Kimura, Yuki Minami},pdfsubject={}, pdfkeywords={}]{hyperref}

\usepackage{amsmath,amssymb,bm}
\usepackage[dvips]{graphicx}
\usepackage{yfonts}
\usepackage{enumerate}

\newcommand{\beq}{\begin{eqnarray}}
\newcommand{\eeq}{\end{eqnarray}}

\newcommand{\U}{\text{U}}
\newcommand{\Lag}{\mathcal{L}}

\newcommand{\delx}{u}
\newcommand{\velocity}{v}

\makeatletter

\@addtoreset{equation}{section}
\makeatother

\begin{document}

\begin{titlepage}

\renewcommand{\thefootnote}{\fnsymbol{footnote}}

\begin{flushright}
RIKEN-QHP-24,
RIKEN-MP-45
\end{flushright}

\vskip5em

\begin{center}
 {\Large {\bf 
 Viscoelastic-electromagnetism and Hall viscosity
 }}

 \vskip3em

 {\sc Yoshimasa Hidaka},\footnote{E-mail address: 
 \href{mailto:hidaka@riken.jp}
 {\tt hidaka@riken.jp}}$^1$
 {\sc Yuji Hirono},\footnote{E-mail address: \href{hirono@nt.phys.s.u-tokyo.ac.jp}
 {\tt hirono@nt.phys.s.u-tokyo.ac.jp}
 }$^{1,2}$
 {\sc Taro Kimura},\footnote{E-mail address: 
 \href{mailto:tkimura@ribf.riken.jp}
 {\tt tkimura@ribf.riken.jp}}$^3$
 and
 {\sc Yuki Minami}\footnote{E-mail address: \href{yminami@riken.jp}
 {\tt yminami@riken.jp}}$^3$
 \vskip2em

{\it 
 $^1$Quantum Hadron Physics Laboratory, RIKEN Nishina Center, Saitama 351-0198,
 Japan \\ \vskip.2em
 $^2$Department of Physics, The University of Tokyo, Tokyo 113-0033, Japan\\ \vskip.2em
 %Hongo 7-3-1, Bunkyo-ku
 $^3$Mathematical Physics Laboratory, RIKEN Nishina Center, Saitama 351-0198,
 Japan 
}

 \vskip3em

 \today
\end{center}

 \vskip2em

\begin{abstract}
%The Hall viscosity current is derived from thermodynamics in one sentence.
We introduce a kind of electromagnetism, which we call 
 viscoelastic-electromagnetism, to investigate viscoelastic transport
 phenomena.
It is shown that Cartan's formalism of general relativity is essential
 for viscoelastic theory, and then the corresponding electric and magnetic
 fields are regarded as a velocity gradient and a Burgers vector density, respectively.
As an application of this formalism, the St\v{r}eda formula for the Hall
 viscosity is obtained.
\end{abstract}

\date{\today}

%\pacs{11.10.Wx, %Finite-temperature field theory
%12.20.-m, %Quantum electrodynamics
%12.38.Mh,	%Quark-gluon plasma 
%52.25.Dg,	 %Plasma kinetic equations
%52.27.Ny	%Relativistic plasmas
%}
%\maketitle

\end{titlepage}

\tableofcontents

\section{Introduction}\label{sec:intro}

According to the linear response theory, we can obtain a number of quantities
characterizing transport phenomena by calculating the correlation
functions in perturbation regime.
In particular, charge transport is the most tractable not only for
theoretical calculation, but also experimental manipulation.
Indeed, the charge current is simply given by introducing the
interaction between matter fields and $\U(1)$ electromagnetic gauge
field, and then differentiating the action with respect to the
gauge potential, $J^\mu = \delta S / \delta A_\mu$.
In this sense, it seems difficult to formulate thermal transport in a
similar manner, because we have to find the corresponding {\em gauge
potential} to be coupled with matter fields.

While we have to involve the charge
conservation, namely the particle number conservation law to investigate
charge transport, we have to deal with the energy conservation in the
case of thermal transport phenomena.
Therefore, it is natural to utilize the framework to describe the energy
itself, namely general relativity.
Luttinger, actually, showed that the gravitational potential is essential
for thermal transport \cite{PhysRev.135.A1505}.
After that it turns out that we can deal with thermal transport in a
quite similar manner by introducing a kind of electromagnetism, called
gravito-electromagnetism \cite{LyndenBell:1996xj}.
In this formalism the gauge potential is coming from a part of degrees
of freedom in the spacetime geometry, and then the thermal gradient can
be represented as the corresponding electric field.
Based on this formalism, various interesting phenomena can be described
by studying the gravitational counterpart of electromagnetism, for
example, the surface state of the topological insulator/superconductor
\cite{Ryu:2010ah,Wang:2010xh,Stone:2012ud}, the St\v{r}eda formula for the thermal Hall
conductivity \cite{Nomura:2011hn}, the thermal analog of the chiral
magnetic effect \cite{Kimura:2011ef}, and so on.

In this paper, we explore a possibility of such an effective electromagnetic
description of viscoelastic theory.
In other words, we have to find the corresponding gauge potential-like
degrees of freedom to this case.
In viscoelastic transport phenomena we essentially investigate transport
of momentum.
Thus, we have to deal with the theory of momentum: it is again general
relativity.
We would like to show that the vielbein, introduced in Cartan's
formalism for general relativity, can be interpreted as such a gauge
potential describing viscoelastic theory, and then investigate some
aspects of the corresponding electromagnetism, which we will call {\em
viscoelastic-electromagnetism}.
In this formalism the $2$-form curvature, the field strength tensor of the
corresponding electromagnetism, is just given by torsion tensor.
Its electric and magnetic components represent the gradient of velocity
and the Burgers vector density, respectively.

Based on this formalism, we then investigate dissipationless
transport in viscoelastic theory.
Such a transport phenomenon is characterized by the Hall viscosity
\cite{Avron:1995fg,springerlink:10.1023/A:1023084404080,Read:2008rn,Kimura:2010yi,Hughes:2011hv,Nicolis:2011ey,Saremi:2011ab,Hoyos:2011ez,Chen:2011fs}.%
\footnote{A similar approach to the Hall viscosity has been quite
recently proposed in \cite{Leigh:2012jv}.}
This quantity is only observed when the time reversal symmetry of the
system is broken, as well as the Hall conductivity.
We will show that, simply generalizing the result for charge transport,
we can obtain the St\v{r}eda formula for the Hall viscosity
\cite{0022-3719-15-22-005,Widom1982474}: the Hall coefficient can be
represented in terms of thermodynamic quantities.

%\vspace{2em}

This paper is organized as follows.
In Sec.~\ref{sec:EM}, we investigate the generic formalism of
electromagnetism with emphasis on its relation to general relativity.
Starting with gravito-electromagnetism as a review, we formulate
viscoelastic-electromagnetism from the vielbein as a fundamental degree of freedom.
We will then clarify the meanings of the electric and magnetic fields, the
gauge transformation, etc, in terms of viscoelastic theory.
In Sec.~\ref{sec:Streda} we derive the St\v{r}eda formula for the Hall
viscosity by applying viscoelastic-electromagnetic formalism.
Introducing the magnetization corresponding to viscoelastic-magnetic
field, we show the Hall viscosity can be expressed in terms of
thermodynamic quantities.
Section~\ref{sec:discussion} is devoted to a summary and discussion.

\if0
References
\begin{itemize}
 \item Original paper by Luttinger \cite{PhysRev.135.A1505},
       introducing gravitational potential
 \item gravito-electromagnetism and thermal Hall effect
       \cite{Ryu:2010ah,Nomura:2011hn} 
 \item Thermal Hall effect via gravitational topological field theory
       \cite{Wang:2010xh}
 \item Hall viscosity 
       \cite{Avron:1995fg,springerlink:10.1023/A:1023084404080,Read:2008rn,Kimura:2010yi,Hughes:2011hv,Nicolis:2011ey,Saremi:2011ab,Hoyos:2011ez}
 \item Chiral heat effect \cite{Kimura:2011ef}
 \item St\v{r}eda formula \cite{0022-3719-15-22-005,Widom1982474}
 \item Chern-Simons-like 3-form for gravito-EM \cite{Volovik:1999wx}
\end{itemize}
\fi

\section{Gravitational electromagnetism}\label{sec:EM}

In this section, we investigate two types of electromagnetisms, which are
coming from general relativity.
One of them is gravito-electromagnetism, used for thermal transport, and
the other is viscoelastic-electromagnetism, for viscoelastic theory.
After reviewing the former one, we will give definitions and detailed
discussions for the latter.

\subsection{Gravito-electromagnetism}\label{sec:gEM}

We would like to show that how gravito-electromagnetism is derived from
general relativity \cite{LyndenBell:1996xj}.
In particular, when the temporal direction is compactified on $S^1$ with
circumference $\beta = 1/T$, we can describe thermal transport by using
such an induced $\U(1)$ gauge field \cite{Ryu:2010ah,Nomura:2011hn,Banerjee:2012iz}.
Thus, this gravito-electromagnetism is regarded as the $\U(1)$ gauge
theory coming from the Kaluza-Klein mechanism.

The $4$-dimensional metric is generally written in the following form,
\begin{equation}
 ds^2 = \gamma_{ij} dx^i dx^j + g_{00} (dx^0 + A_i dx^i)^2,
\end{equation}
where we parametrize 
\begin{equation}
 A_i = \frac{1}{g_{00}} g_{0i},\quad \gamma_{ij}= g_{ij} - \frac{g_{0i}g_{0j}}{g_{00}} .
  \label{gravito_vector_pot}
\end{equation}
We will see that gravito-electromagnetism is formulated based on this
$3$-dimensional $1$-form as the $\U(1)$ gauge field.
Then we introduce the following parametrization,
\begin{equation}
 g_{00} = -e^{2\sigma}. %= -\frac{T_0}{T}, 
\end{equation}
This $\sigma$ is called {\em dilaton}, which plays a role as the scalar
potential for gravito-electromagnetism.
In fact, when the temporal direction is compactified on $S^1$ with
circumference $\beta=1/T$, this part of the metric is proportional to
$\beta^2$, so that the temperature dependence of the dilaton yields
$\sigma =- \log (T/T_0)$, where $T_0$ is the temperature at thermal equilibrium. 
Thus, the thermal gradient is given by
\begin{equation}
 \nabla_i \sigma = - \frac{\nabla_i T}{T}.
 %{E}^{ia} = F^{i0}= \partial^i A^0 - \partial^0 A^i
  \label{dilaton_gradient}
\end{equation}
Since the right hand side of this equation is just regarded as the
electric field of gravito-electromagnetism, $E_i = - \nabla_i T / T$,
the dilaton is nothing but the scalar potential for gravito-electromagnetism.
Note that this gravito-electric field is also represented as the
temporal derivative of the vector potential in Eq.~(\ref{gravito_vector_pot}),
and the gravito-magnetic field is similarly introduced as 
\begin{equation}
B^i=\epsilon^{ijk} \partial_i A_j,
\end{equation}
where $\epsilon^{ijk}\equiv\epsilon^{0ijk} e^{-\sigma}$ with $\epsilon^{0123}=1/\sqrt{-g}$.
This quantity is essentially related to the rotation of the system
\cite{LyndenBell:1996xj}.
Indeed the conjugate quantity to this magnetic field, namely the
corresponding magnetization, yields the angular-momentum operator
\cite{Nomura:2011hn},
\begin{equation}
 M_i = \epsilon_{ijk} x^j T^{k0} = L_i,
\end{equation}
where $T^{\mu\nu}$ stands for the stress tensor.

The gauge transformation for this $\U(1)$ connection is coming from the
higher-dimensional coordinate group $x'^0 = x^0 + \lambda(x^i)$,
\begin{equation}
 A'_i = A_i - \partial_i \lambda.
\end{equation}
Thus, we can define the conserved current corresponding to this $\U(1)$
symmetry.
Since this is coming from the conservation law of energy, we can deal
with thermal transport with this formalism.

\subsection{Viscoelastic-electromagnetism}\label{sec:vEM}

In this section, we formulate another kind of electromagnetism, {\em
viscoelastic-electromagnetism}, in order to deal with viscoelastic
theory from the field theoretical point of view.
Viscoelastic transport is based on momentum conservation, while electric
and thermal ones are coming from carrier and energy conservations, respectively.
Therefore, we have to introduce degrees of freedom describing
momentum transport with manifesting its conservation.
We would like to show that {\em vielbein} plays a fundamental role in
constructing viscoelastic-electromagnetism.

Let us then start with the vielbein formalism of the curved spacetime,
which is also called {\em Cartan's formalism}.
The metric of the spacetime is generally written by using
vielbein $e_\mu^{~a}$, which connects the global curved spacetime coordinate
with the locally flat coordinate,
\begin{equation}
 g_{\mu\nu} = \eta_{ab} \, e_\mu^{~a} e_\nu^{~b} ,
\end{equation}
where $\eta_{ab}=\textrm{diag}(-1,1,1,1)$ is the locally flat Minkowski metric.
In this formalism each coordinate is dealt with separately.
Thus, this vielbein can be regarded as a $1$-form with respect to the global
curved spacetime coordinate.

We can provide a physical meaning for the vielbein in viscoelastic theory.
The time component of the vielbein $e_\mu^{~0}$ coincides with 
the gauge fields of gravito-electromagnetic fields $e_\mu^{~0}=(e^\sigma, e^\sigma A_i)$ discussed in the previous section.
For the spatial component,  introducing the displacement vector $\delx^i$, which is regarded as 0-form,
the vielbein can be expanded as $e_\mu^{~i} = \delta_\mu^{~i} +
\partial_\mu \delx^i$ in the linear order \cite{ChaikinLubensky}.
Here $\partial_\mu \delx^i \equiv \delx_\mu^{~i}$ is the distortion tensor,
which is just $1$-form with respect to the global coordinate.
Thus, the vielbein plays almost the same role as this distortion tensor.
Therefore, we define electromagnetism corresponding to this $1$-form as {\em viscoelastic-electromagnetism}.
The $2$-form, obtained from the vielbein, yields the torsion tensor,
\begin{equation}
 T^a = d e^a + \omega^{a}_{~b} \wedge e^b .
\end{equation}
In the following we assume the spin connection is zero for simplicity.
Thus, the field strength $2$-form is given by
\begin{equation}
 F_{\mu\nu}^{~~a} = \partial_\mu e_\nu^{~a} - \partial_\nu e_\mu^{~a} .
\end{equation}
If this $2$-form takes non-zero value, the displacement vector, namely the
vector potential for viscoelastic-electromagnetism, cannot be
single-valued.
Actually, the corresponding magnetic field is given by
\begin{equation}
 B^{ia} = \epsilon^{ijk} \partial_j e_{k}^{~a},
  \label{visco-magnetic_field}
\end{equation}
and the magnetic flux 
\begin{equation}
b^a=\int d\vec{S}\cdot \vec{B}^a,
\end{equation}
is identified as the {\em Burgers vector}, which characterizes a lattice
dislocation in a crystal.
Similarly, the electric field yields Eq.~(\ref{dilaton_gradient}) and
\begin{equation}
 E_i^{~j} \equiv F_{0i}^{~~j} = -\partial_i\,e_0^{~j} \equiv -\partial_i \velocity^{j},
 \label{eq:Eij}
\end{equation}
where we defined the spatial derivative of the vielbein as the gradient
of the velocity field due to the relation between the vielbein and the
distortion fields. 
This shows the velocity can be interpreted as the chemical potential for
viscoelastic-electromagnetism, and also that we can formulate the momentum
transport response by applying this viscoelastic-electric field.

We can show that the gauge transformation is consistent with
viscoelastic theory.
Taking the transformation for the $1$-form,
$e_\mu^{~a} \to e_\mu^{~a} + \partial_\mu w^a$, it does not affect the
torsion tensor $2$-form because we have $\epsilon^{\mu\nu} \partial_\mu
\partial_\nu w^a = 0$.
This symmetry is coming from the conservation of energy and momentum, i.e., translation symmetry,
 %{\bf(conservation law: $\partial^\mu T_{\mu\nu}^{~~a}=0$~?)}
and thus, we can derive the corresponding momentum current as N\"other's current.
% On the other hand, this corresponds to the curved spacetime in the
% presence of the torsion cannot be described as Riemann spacetime.

Let us comment on the topological action, which plays an important role
in dissipationless transport.
The $(2+1)$-dimensional topological action for viscoelastic theory is
constructed in a similar way to the Chern-Simons action \cite{Hughes:2011hv},
\begin{equation}
 S = \frac{1}{32\pi} \frac{2}{\ell^2} \int d^3 x \, (\det e)
 \eta_{ab} \epsilon^{\mu\nu\rho} \,   e_\mu^{~a} \partial_\nu e_{\rho}^{~b}.
  \label{visco_CS_action}
\end{equation}
The covariant form of this action is given by $S \sim \int e^a \wedge T_a$.
Note that there exists a dimensionful constant $\ell$ in this topological action.
This factor makes the topological origin of this action
ambiguous.\footnote{See, for example,
\cite{Chandia:1997hu,Kreimer:1999yp,Chandia:1999az,Li:1999ue}.}
The energy-momentum tensor reads
\begin{equation}
T^{\mu}_{~a}= \frac{\hbar}{8\pi\ell^2}\epsilon^{\mu\nu\rho}\partial_\nu e_{\rho a}.
\label{eq:enegyMomentumTensor}
\end{equation}
The viscoelastic transport coefficient, the Hall viscosity, is simply
obtained from the corresponding Chern-Simons action (\ref{visco_CS_action}),
\begin{equation}
 \eta_{\rm H} = \frac{\hbar}{8\pi\ell^2} .
\end{equation}
Identifying the dimensionful constant as the magnetic length $1/\ell_{\rm
B}=\sqrt{eB/\hbar}$, this reproduces well-known result.
The contribution of the energy current to the thermal Hall conductivity 
can be obtained from Eq.~(\ref{eq:enegyMomentumTensor}).
In general, the thermal-conductivity tensor is defined as $Q^i = T^{i}_{~0} - \mu J^i=-\kappa^{ij}\partial_j T$, where $Q^i$, $\mu$, and $J^i$ are the heat current,  the chemical potential, and the charge current, respectively.
Therefore,  the contribution from $T^i_{~0}$ to the thermal Hall conductivity is 
${\hbar T}/(8\pi\ell^2)$, which is proportional to the Hall viscosity.
%\begin{equation}
%S= \int d^4x \eta_{ab}e^{\mu\nu\rho\sigma}e^a_\mu e^b_\nu  \partial_\rho A_\sigma
%\end{equation}

Similarly, the $(3+1)$-dimensional topological action, the viscoelastic analog
of the $\theta$-term, is given by
\begin{equation}
 S = \frac{1}{32\pi^2} \frac{2}{\ell^2} \int d^4 x \, (\det{e})\, \theta(x) \,
\eta_{ab}  \epsilon^{\mu\nu\rho\sigma} \, \partial_\mu e_{\nu}^{~a} \partial_\rho e_{\sigma}^{~b} .
\end{equation}
The covariant form of this action is directly related to Nieh-Yan term,
$T^a \wedge T_a - R_{ab} \wedge e^a \wedge e^b$
\cite{Nieh:1981ww}.
In terms of viscoelastic-electromagnetism the Lagrangian is written as
an inner product of electric and magnetic fields,
\begin{equation}
 S = \frac{1}{32\pi^2} \frac{2}{\ell^2} \int d^4  x \, (\det{e}) \, \theta(x) \,
\eta_{ab}  \vec{E}^a \cdot \vec{B}^b.
  \label{VE_theta_term}
\end{equation}
When the $\theta$-angle has spatial dependence, we observe the Hall
viscosity on the domain-wall.
In particular, in the time reversal symmetric system, the $\theta$-angle
has to be $\theta = 0$ or $\pi$.
Therefore, this $\theta$-term is reduced to the Chern-Simons action
(\ref{visco_CS_action}) on the boundary of the topological insulator
\cite{Hasan:2010xy,Hasan:2010hm}
because the gradient of the $\theta$-angle is
proportional to $\delta$-function.
On the other hand, applying the temporal dependence of the
$\theta$-angle, namely $\dot{\theta}\not=0$,
the momentum current is proportional to viscoelastic-magnetic
field as $J^i_{~a} \, (=T^i_{~a}) \sim (\dot{\theta}/\ell^2) B^{i}_{~a}$.
This is just the viscoelastic analog of the chiral magnetic effect
\cite{Kimura:2011ef}.

\section{St\v{r}eda formula for Hall viscosity}\label{sec:Streda}

In this section, we derive the St\v{r}eda formula for the generic
electromagnetic formalism.
In the case of the usual electromagnetism, this yields the
formula expressing the Hall conductivity in terms of bulk quantities
\cite{0022-3719-15-22-005,Widom1982474}.
Recently, it has been shown that this formula is also possible for the
thermal Hall system \cite{Nomura:2011hn}.
We would like to show that such a useful formula for the Hall viscosity
is obtained from viscoelastic-electromagnetism.

We now start with the standard electromagnetism.
Let us introduce a magnetization, which is conjugated to the
corresponding magnetic field,
\begin{equation}
 M_i = - \frac{\delta S} {\delta B^i}
  = - \frac{\delta S}{\delta A_j} \frac{\partial A_j}{\partial B^i}.
  \label{magnetization01}
\end{equation}
Recalling the derivative of Lagrangian with respect to the gauge field
gives rise to the conserved current, and substituting the
expression of the symmetric gauge in the presence of magnetic field,
$A_i = -\epsilon_{ijk} x^j B^k/2$, the magnetization
(\ref{magnetization01}) is rewritten as
\begin{equation}
 M_i = \frac{1}{2}\epsilon_{ijk} x^j J^k .
\end{equation}
The Hall current is given by using the magnetization:
\begin{equation}
 J_{\rm H}^i = \epsilon^{ijk} \partial_j M_k . 
\end{equation}
In order to study the conductivity, we should rewrite this
expression in terms of the corresponding electric field.
Since the gradient of the chemical potential yields the electric field, $E_i=\partial_i\mu/e$,
we have the relation between the current and the electric field,
\begin{equation}
 J_{\rm H}^i = \epsilon^{ijk} 
  \left(\frac{\partial M_k}{\partial \mu}\right)_{T,B} \partial_j \mu
  = \epsilon^{ijk} \left(\frac{\partial M_k}{\partial \mu}\right)_{T, B} eE_j, %E_i = \partial_i \mu
\end{equation}
where $e$ is the electric charge.
This means the corresponding Hall transport coefficient is given by
\begin{equation}
 \sigma_{\rm H} = e\left(\frac{\partial M_z}{\partial \mu}\right)_{T, B} .
\end{equation}
The conjugate quantities to the magnetization and the chemical
potential are magnetic field $B$ and particle number density $n$, respectively.
Thus, by using the thermodynamic relation, as discussed in
\cite{Widom1982474},
this Hall coefficient is also written as
\begin{equation}
 \sigma_{\rm H} = e\left(\frac{\partial n}{\partial B}\right)_{T, \mu} .
\end{equation}
This is so-called the St\v{r}eda formula for the Hall conductivity
\cite{0022-3719-15-22-005,Widom1982474}.
While this original formula is of the electric Hall current, its
generalization to the thermal transport is also proposed by applying
essentially the same procedure to gravito-electromagnetism
\cite{Nomura:2011hn}.

Let us then apply this derivation to viscoelastic-electromagnetism based
on Sec.~\ref{sec:vEM}.
The corresponding magnetization is similarly introduced as
\begin{equation}
 M_{ia} 
  = - \frac{1}{\det e}\frac{\delta S}{\delta B^{ia}}
  = - \frac{1}{\det e}\frac{\delta S}{\delta e_j^{~b}}
  \frac{\partial e_j^{~b}}{\partial B^{ia}} .
\end{equation}
We now define the current as the derivative of Lagrangian as well as the
other electromagnetisms,
\begin{equation}
 {J^i}_a =  \frac{1}{\det e}\frac{\delta S}{\delta e_i^{~a}}=
 \frac{\partial \Lag}{\partial e_i^{~a}} +{e^i}_a\Lag,
\end{equation}
where $\Lag$ is the Lagrangian density.
Thus, the corresponding current to viscoelastic theory turns out to
be the spatial part of the stress tensor, including both of indices for
local and global coordinates, ${J^i}_a = {T^i}_{a}$.
Substituting the symmetric gauge configuration to this expression,
$e_i^{~a} = \epsilon_{ijk} x^j B^{ka}$, the magnetization is rewritten as
\begin{equation}
 M_{ia} = \frac{1}{2}\epsilon_{ijk} x^j {T^k}_a .
\end{equation}
This is quite analogous to the gravito-magnetization, which is given by
the angular momentum operator, $M_i = \epsilon_{ijk} x^j T^{k0}/2 =	L_i$.
Then, the Hall current in terms of the magnetization is given by 
\begin{equation}
 {J_{\rm H}^i}_a
  = \epsilon^{ijk} \partial_j M_{ka} .
\end{equation}
Then, applying the chain rule, we have the following expression for the
stress tensor,
\begin{equation}
 {J^i_{\rm H}}_a = \epsilon^{ijk}
  \left( \frac{\partial M_{ka}}{\partial v_b} \right)_{T, B} \partial_j \velocity_b
  = \epsilon^{ijk}
    \left( \frac{\partial M_{ka}}{\partial v_b} \right)_{T, B} E_{jb},
\end{equation}
where $\velocity_b=e^0_{~b}$ is the fluid velocity  introduced in Eq.~(\ref{eq:Eij}).
Since the coefficient of the gradient of velocity in the stress tensor
gives the viscosity tensor, we obtain 
\begin{equation}
% \eta_{i~j}^{~a~b} 
% {{\eta^{ij}}_a}^b
 \eta^{ij~b}_{~~a}
  = \epsilon^{ijk} 
  \left( \frac{\partial M_{ka}}{\partial v_b} \right)_{T, B}.
  \label{visco_tensor}
\end{equation}
The shear viscosity is symmetric under any exchange of indices.
However, this viscosity tensor includes anti-symmetric part, giving rise to
dissipationless transport of momentum.
The Hall viscosity, which characterizes such a dissipationless
transport, is given by the anti-symmetric part of the viscosity tensor (\ref{visco_tensor}).
Therefore, we obtain the St\v{r}eda formula for the Hall viscosity,
\begin{equation}
 {\eta_{{\rm H}, a}}^b
  = \left( \frac{\partial M_{a}}{\partial v_b} \right)_{T, B}
  = \left( \frac{\partial P^b}{\partial B^{a}} \right)_{T, v} .
\end{equation}
Note that velocity $v^b$ is conjugate to momentum $P^b$.
Usually this part should be diagonal by assuming the isotropic
configuration, ${\eta_{{\rm H}, a}}^b \propto {\delta_{a}}^b$.

This St\v{r}eda formula can be generalized to $(3+1)$-dimensional case
to obtain the magnetoelectric polarizability, which characterizes the
cross-response between electric and magnetic fields \cite{Nomura:2011hn}.
In analogy with this we can introduce the analogous quantity,
characterizing the relation between viscoelastic-electromagnetic fields, 
viscoelastic-magnetoelectric polarizability,
\begin{equation}
 \chi_{ia}^{jb} 
  = \frac{\partial M_{ia}}{\partial E^{jb}}
  = \frac{\partial P_{jb}}{\partial B^{ia}} ,
\end{equation}
where $P_{ia}$ stands for the conjugate variable to viscoelastic-electric
field.
It is quite natural that this quantity describes the cross-correlation
between electric and magnetic responses because it is also written as $\chi = -
\partial^2 \Lag / \partial E \partial B$.

\section{Summary and discussion}\label{sec:discussion}

In this paper, we have introduced the idea of
viscoelastic-electromagnetism as a useful method to deal with
viscoelastic theory.
We have then obtained the St\v{r}eda formula for the Hall viscosity as an
application of such a generalized electromagnetism.

We have shown that the vielbein, which is used in Cartan's
formalism of general relativity, can be interpreted as the vector
potential for viscoelastic theory.
The corresponding electric and magnetic fields have natural meanings in
terms of viscoelastic theory, velocity gradient and the Burgers vector,
respectively.
We have pointed out that the gauge symmetry for this formalism is also
reasonable in viscoelastic theory, and the topological action is
constructed quite similarly to the usual electromagnetism.

We have then derived the St\v{r}eda formula for the Hall viscosity.
We can follow almost the same procedure to obtain this formula by using
viscoelastic-electromagnetic formalism.
We have clarified the viscoelastic analog of the magnetization, which is
essential for the derivation of the St\v{r}eda formula
Then, generalizing this result to $3$-dimensional theory, we have obtained
viscoelastic-magnetoelectric polarizability, which
characterizes the cross-correlation between viscoelastic-electric and
magnetic responses. 

Let us comment on possibilities of future works along this direction.
First it is quite interesting to clarify the relation between the Hall
conductivity and the Hall viscosity in terms of
viscoelastic-electromagnetism \cite{Hoyos:2011ez}.
It is shown that the cross-correlation term, called Wen-Zee term
\cite{Wen:1992ej}, plays an essential role in the derivation of such a
relation.
Thus, we have to somehow involve such an interesting term in
electromagnetic formalism.
It seems also important to search any other cross-responses between those
Hall transport coefficients.
Second possibility is extension of the St\v{r}eda formula to other
dissipationless transport.
Indeed, the St\v{r}eda formula for spin Hall conductivity is investigated in
\cite{PhysRevB.73.073304}.
Thus, it is interesting to apply it to spin Hall version of viscosity,
which is discussed in \cite{Kimura:2010yi}, as a simple generalization
of the formula derived in this paper.
Finally, it is possible to apply viscoelastic-electromagnetic formalism to
holographic description of the strongly correlated systems, which is
based on AdS/CFT.
In particular, it is meaningful to explore its stringy origin by
considering D-brane construction.

\subsection*{Acknowledgements}

The authors thank  A.~Furusaki, T.~Nishioka and N.~Ogawa.
The research of Y.~Hidaka is  supported by a Grant-in-Aid for Scientific Research (No.23340067, 24740184) from the Ministry of Education, Culture, Sports, Science and Technology (MEXT) of Japan;
of T.~Kimura, by Grant-in-Aid for JSPS Fellows (No.~23-593);
of Y.~Hirono,  by Grant-in-Aid for JSPS Fellows (No.~23-8694).
%%%%%%% References %%%%%%%

\bibliographystyle{ytphys}
\bibliography{Hall}
 
\end{document}